\documentstyle[11pt,newpasp,twoside,epsf]{article}
\begin{document}
\title{VLBA Calibrator Survey: Astrometric and Image Results}
  \author{Leonid Petrov}
  \affil{NVI, Inc./NASA GSFC, Code 926, Greenbelt, 20771 MD, USA}
  \author{David Grodon}
  \affil{Raytheon ITSS/NASA GSFC, Code 926, Greenbelt, 20771 MD, USA}
\author{Anthony Beasley}
\affil{Owens Valley Radio Observatory, California Institute of Technology,
       P.O. Box 968, Big Pine, CA 93513, USA}
\author{Ed Fomalont}
\affil{National Radio Astronomy Observatory, Edgemont Rd., Charlottesville,
VA 22903, USA}
 
\begin{abstract}
   Positions and maps of 1608 new compact sources were obtained in twelve
sessions observed during 1994--2002 at the VLBA network at 8.4/2.3 GHz.
These sources are recommended for use as calibrators for phase reference
imaging and as geodetic sources for astrometric/geodetic VLBI applications.
\end{abstract}
 
\section{Introduction}
  The purpose of the VLBA Calibrator Survey 
(Beasley et al. 2002), (Fomalont et al. 2003) was to extend the list of compact
sources with coordinates known at the milliarcseconds level for use as
calibrators for mapping and as target sources for geodetic applications.
In total, the positions of 1608 new sources were obtained. As a result of
processing VCS sessions the total number of sources with precise coordinates
increased by a factor of three. The goals of the survey were: a) to increase
the surface density of known geodetic-grade calibrators with mas-accurate
positions; b) to facilitate routine phase-referencing allowing high-resolution
radio imaging of weak target sources; c) to provide  a uniform image database
at 2.3 and 8.4 GHz for use in scientific applications.
 
\section{Observations}
 
   Observations were carried out in twelve 24 hour sessions. Observations used
the VLBA dual-frequency geodetic mode observing simultaneously at 2.3 and
8.4 GHz. Signal was recorded at four 8 MHz wide baseband channels over a large
spanned bandwidth (100 MHz at S-band and 400 MHz at X-band). Each source was
observed 2--3 times for 60--90 sec.

\section{Astrometric catalogue}
 
\begin{figure}
   \par\vspace{-2mm}\par
   \epsfclipon \plotone{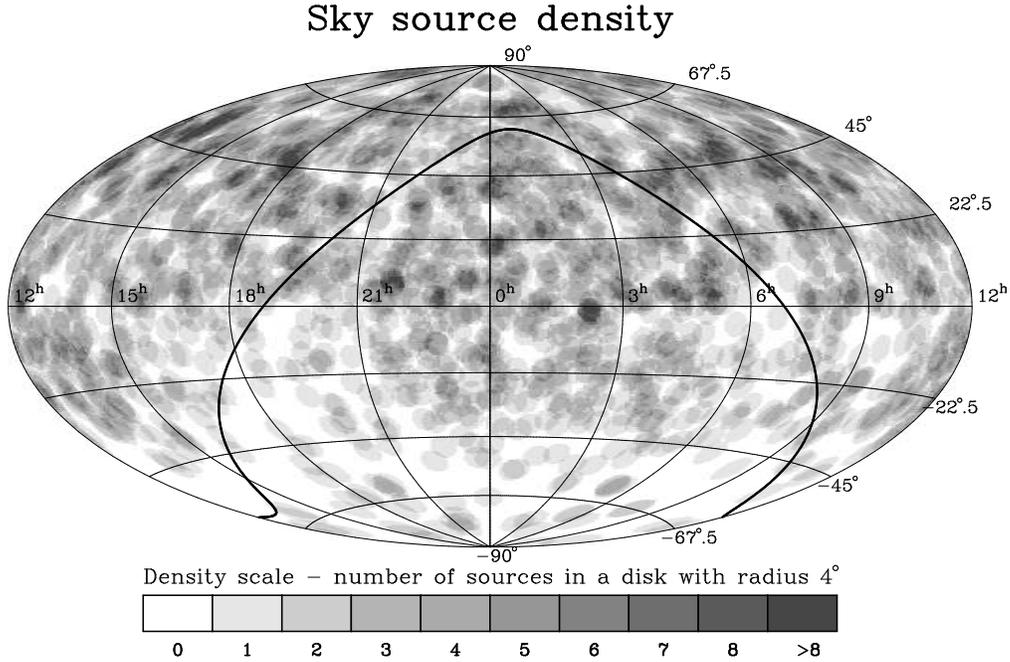}
   \caption{Sky source density. Solid line is the galactic plane.}
   \par\vspace{-2mm}\par
   \label{f:f1}
\end{figure}

   Before analysis of the VCS sessions the number of sources observed under
geodesy/astrometry programs with positions known better than 1 nrad was 615.
After analysis of the VCS campaign this number increased to 2011. Now there is
at least one calibrator within $4^\circ$ of any target direction at 90.1\% of
the sky at declinations $\delta > -40^\circ$, as shown in figure~\ref{f:f1}.
 
  Some 15\% of sources from the VCS lists were observed previously in
astrometric/geodetic programs. Analysis of the differences between the VCS
catalogues and Goddard gsf2003c catalogue allowed us to assess the error model.
As a result of this analysis the formal uncertainties were inflated in this
way:
\begin{equation}
  \sigma(\alpha)_r = \sqrt{ (1.5 \sigma_\alpha )^2 + E(\alpha)^2/cos^2\delta}
  \qquad\qquad
  \sigma(\delta)_r = \sqrt{ (1.5 \sigma_\delta )^2 + E(\delta)^2}
\end{equation}
 
$E(\alpha)$ and $E(\delta)$ are declination dependent:
 
\begin{table}[h]
  \vspace{-2ex}
  \begin{center}
    \begin{tabular}{|l l l|}
      \tableline
          Dec Zone                   & $   E(\alpha) $  & $ E(\delta) $  \\
          (deg)                      &     (mas)        &  (mas)         \\
      \tableline
         $ [+20^\circ, +90^\circ] $  &     0.3          &  0.3           \\
         $ (-30^\circ, +20^\circ] $  &     0.3          &  0.6           \\
         $ (-46^\circ, -30^\circ] $  &     0.6          &  1.1           \\
      \tableline
     \end{tabular}
  \end{center}
  \vspace{-3ex}
\end{table}
 
\section{Image results}
 
  Amplitude and initial phase calibration was done by AIPS software.
Sources were imaged by the Caltech DIFMAP package. Some 70\% of sources were
imaged automatically, the others were imaged manually. The typical image rms
is 2--3 mJy, with dynamic range 30:1 or better.
 
\begin{figure}[t]
   \mbox{
         \epsfxsize=50mm \epsffile{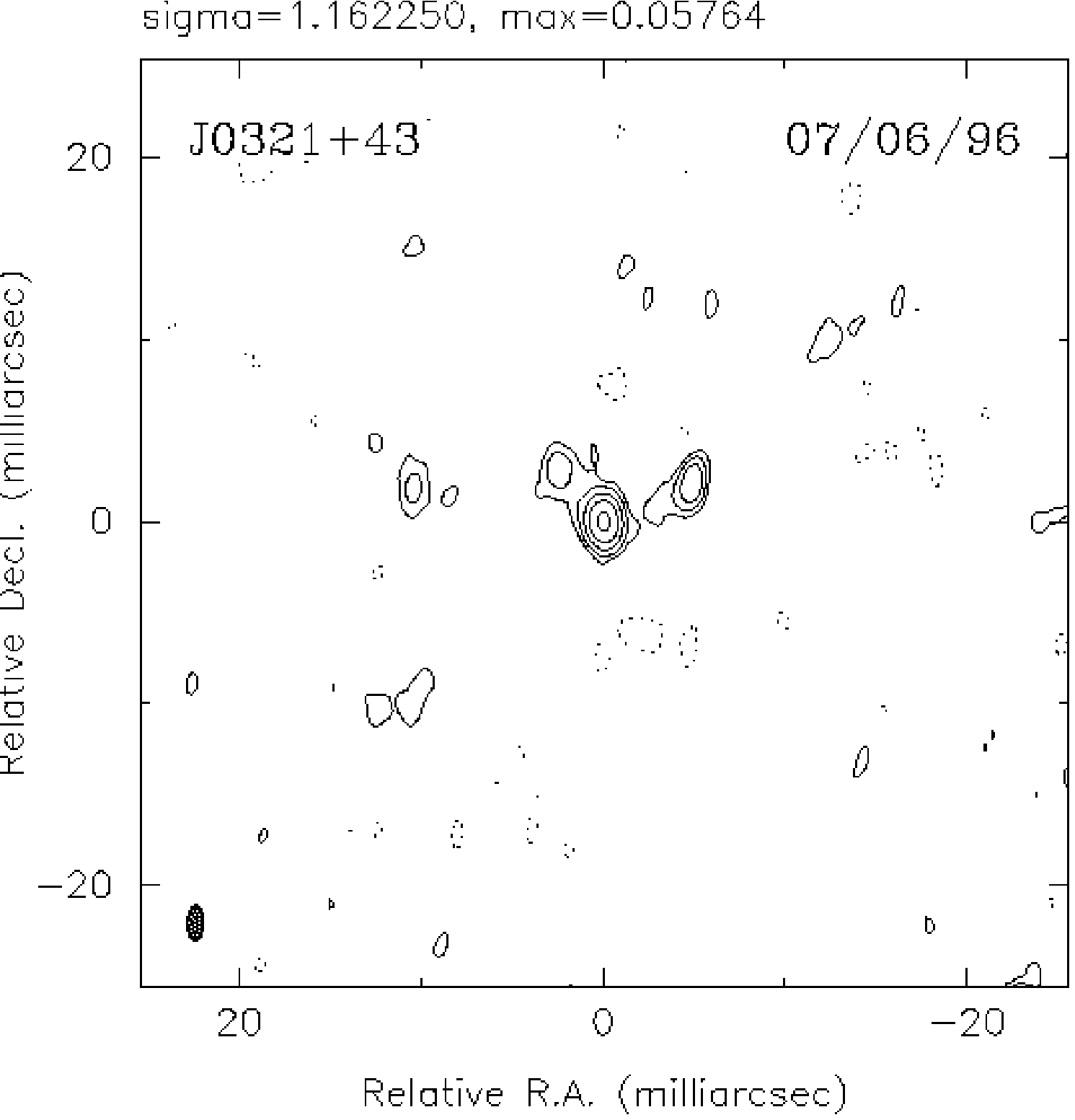} \hspace{10mm}
         \epsfxsize=50mm \epsffile{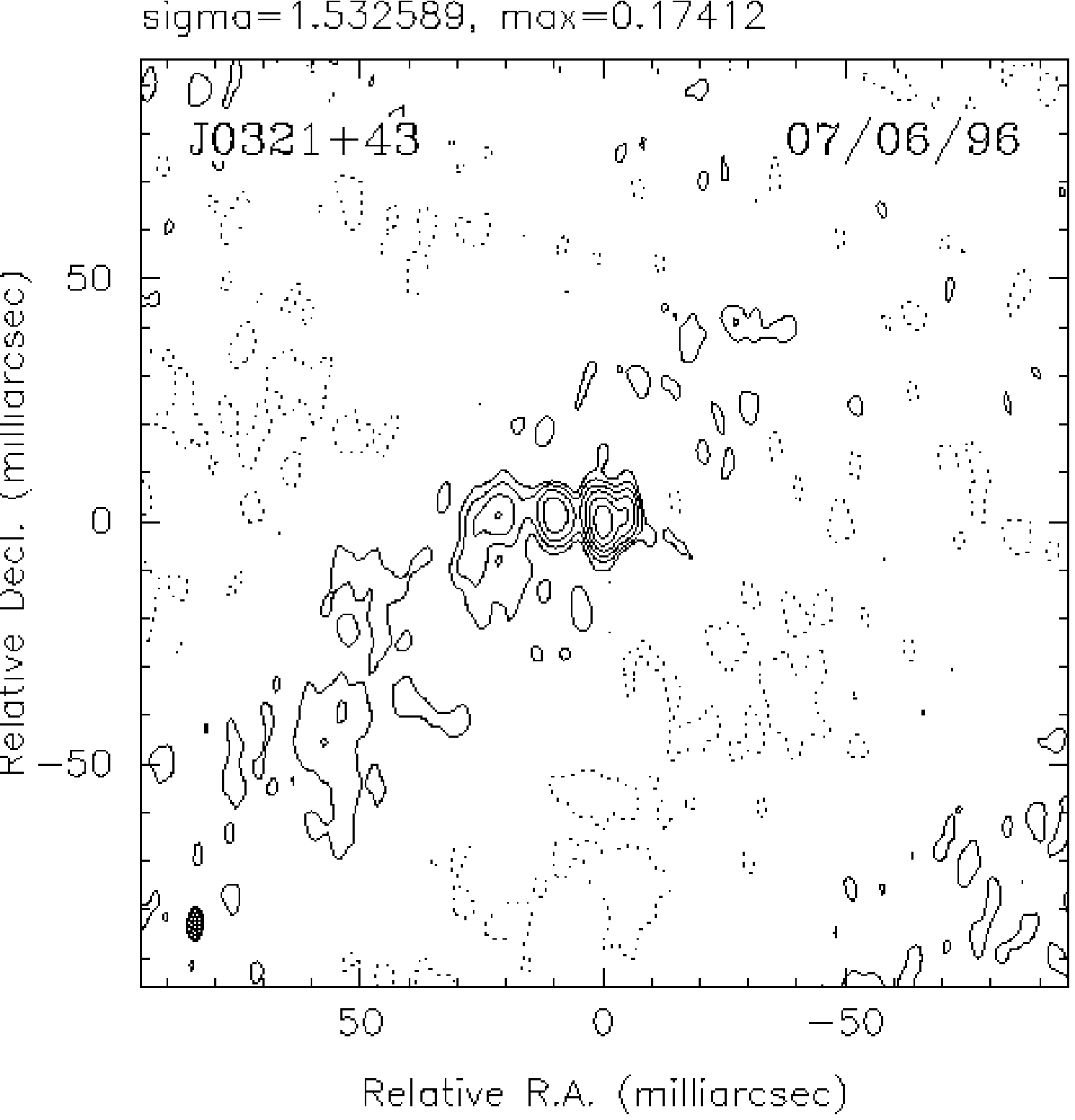}
        }
   \epsfclipon
   \par\vspace{4mm}\par
   \mbox{
         \epsfxsize=50mm \epsffile{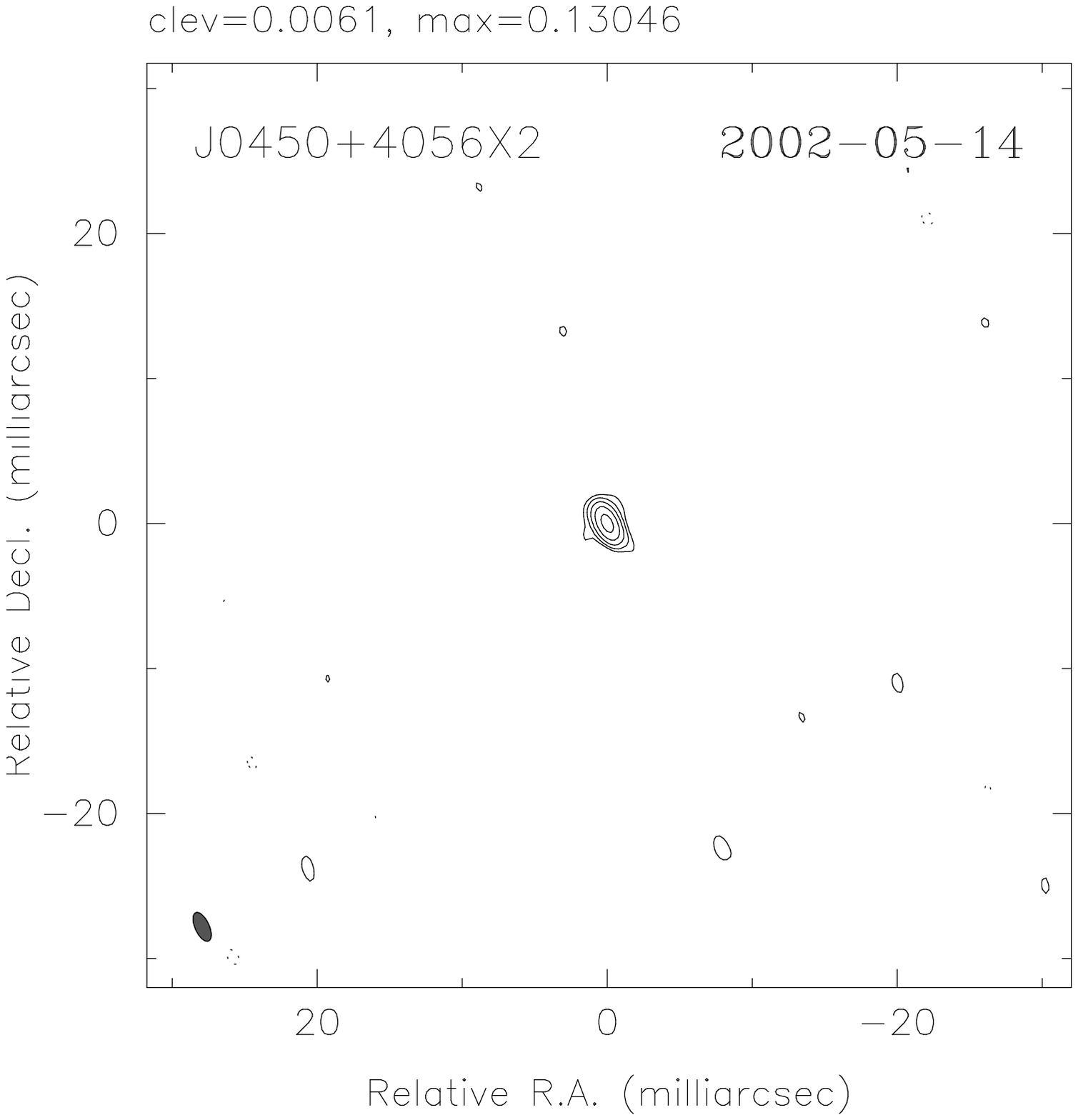} \hspace{10mm}
         \epsfxsize=50mm \epsffile{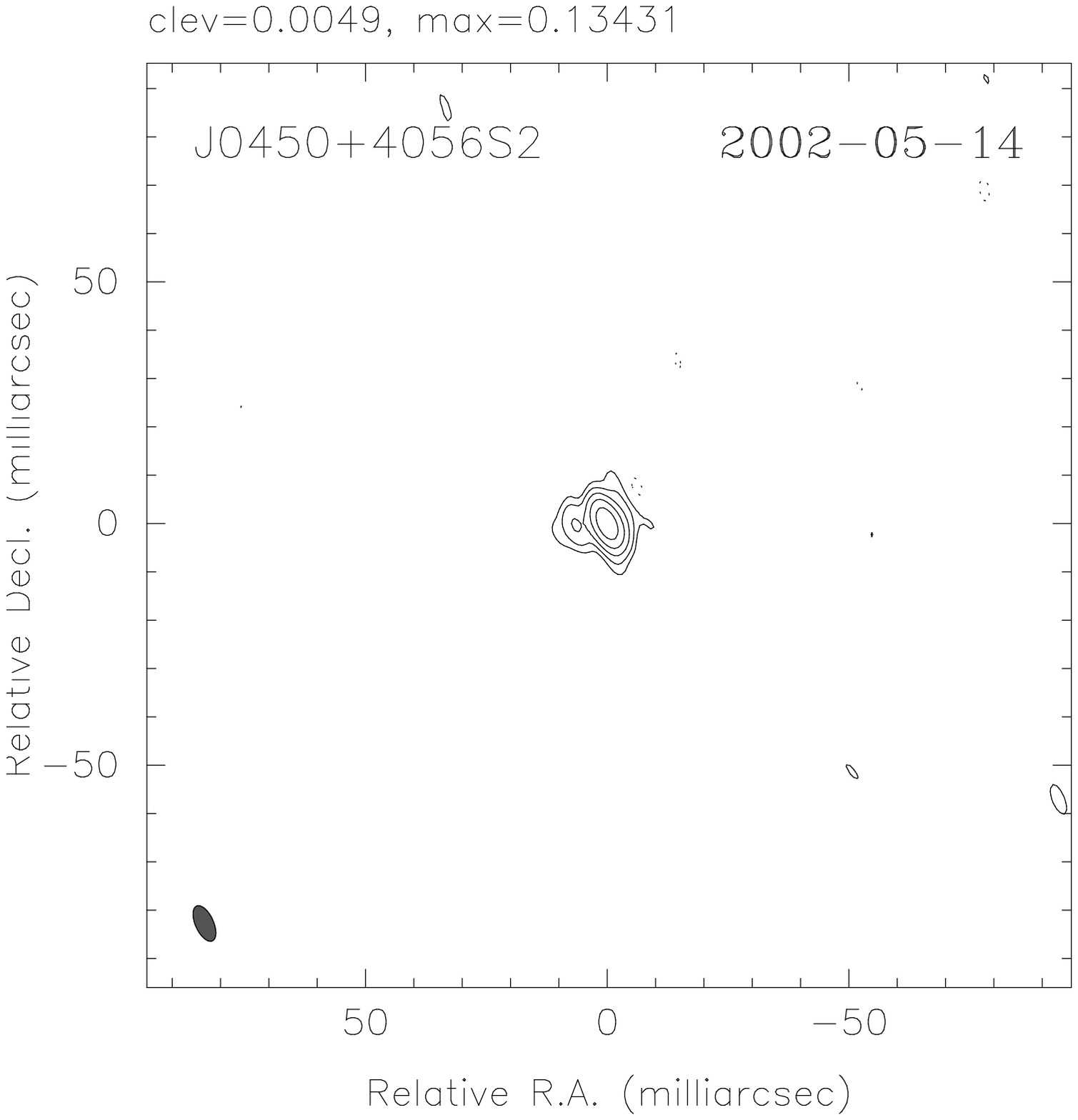}
        }
   \caption{Example of images at X-band (left column) and S-band (right column)}
   \label{f:f2}
\end{figure}
 
  The image database and a source search engine are available on the Web at
{\tt http://magnolia.vlba.nrao.edu/vlba\_calib} . The astrometric catalogue
is in {\tt http://gemini.gsfc.nasa.gov/solutions/astro} .


\begin{references}
\reference
     {Beasley,~A.J., Gordon,~D., Peck,~A.B., Petrov,~L., MacMillan,~D.S.,
         Fomalont~E.B., Ma,~C., 2002, \apjs, 141, 13.}
 
\reference
    {Fomalont~E., Petrov~L., McMillan~D.S., Gordon~D., Ma~C., 2003,
         submitted to \aj.}
\end{references}
\end{document}